# Unusual compressibility in the negative-thermal-expansion material $ZrW_2O_8$

C. Pantea,[1] A. Migliori,[1] P. B. Littlewood,[2,3] Y. Zhao,[4] H. Ledbetter,[1] T. Kimura,[5] J. Van Duijn,[6] G. R. Kowach[7,†]

[1]*National High Magnetic Field Laboratory, Pulsed Field Facility, LANL, Los Alamos, NM 87545, USA.* [2]*Cavendish Laboratory, University of Cambridge, Madingley Road, Cambridge CB3 OHE, UK.* [3]*National High Magnetic Field Laboratory, Tallahassee, FL.* [4]*LANSCE Division, LANL, Los Alamos, NM 87545, USA.* [5]*Condensed Matter & Thermal Physics, LANL, Los Alamos, NM 87545, USA.* [6]*Department of Physics and Astronomy, Johns Hopkins University, Baltimore, MD 21218, USA.* [7]*Lucent Technologies/Bell Laboratories, Murray Hill, New Jersey 07974, USA.*

The negative thermal expansion (NTE) compound $ZrW_2O_8$ has been well-studied because it remains cubic with a nearly constant, isotropic NTE coefficient over a broad temperature range.[1] However, its elastic constants seem just as strange as its volume because NTE makes temperature acts as *positive* pressure, decreasing volume on warming and, unlike most materials, the thermally-compressed solid *softens*[2]. Does $ZrW_2O_8$ also soften when pressure alone is applied? Using pulse-echo ultrasound in a hydrostatic SiC anvil cell, we determine the elastic tensor of monocrystalline $ZrW_2O_8$ near 300 K as a function of pressure. We indeed find an unusual decrease in bulk modulus with pressure. Our results are inconsistent with conventional lattice dynamics, but do show that the thermodynamically-complete constrained-lattice model[3] can relate NTE to elastic softening as increases in either temperature or pressure reduce volume, establishing the predictive power of the model, and making it an important concept in condensed-matter physics.

---

[†] Present address: The City College of New York, The City University of New York, Department of Chemistry, 138th Street and Convent Avenue, New York, NY 10031.

The negative-thermal-expansion (NTE) compound $ZrW_2O_8$ has been well studied because from 50 K to 1050 K it remains cubic with a nearly constant NTE coefficient.[1] However, its elastic constants seem just as strange as its volume because like most solids, its elastic constants soften on warming.[2] But because of the NTE, warming decreases volume just as pressure would, and most solids *stiffen* when compressed. Using pulse-echo ultrasound in a hydrostatic SiC anvil cell,[4] we have determined the complete elastic tensor of monocrystalline $ZrW_2O_8$ near 300 K as a function of pressure and found that $ZrW_2O_8$ softens when compressed, then undergoes a phase transition at 0.5 GPa, a much higher pressure than previously reported for powdered material.

The thermodynamically-complete framework-solid[3] model for NTE is substantially different from earlier attempts at such a picture because it provides many degrees of freedom, instead of only one. This makes it applicable in only those complex structures where some unconstrained "open space" exists, such as $ZrW_2O_8$, and perhaps molecules important to biophysics such as proteins. In some other phase transitions, such as in $La_2CuO_4$[5] or $Lu_5Ir_4Si_{10}$,[6] effects of the phase transition extend both well above and well below the critical temperature or pressure, but it is the open –but-constrained lattice structure, lacking in these examples, not the phase transition itself, that appears to induce unusual behaviour $ZrW_2O_8$. Thus it is the constrained-lattice-induced NTE (not the pressure-induced phase[7]) that overwhelms the ordinary positive expansion and it is the constrained lattice itself that provides the mechanism for elastic moduli to soften as volume decreases, whether such decrease is induced by pressure, or rising temperature.

The pressure-induced phase transition from α-ZrW$_2$O$_8$ (cubic) to γ-ZrW$_2$O$_8$ (orthorhombic) was determined previously in powder samples[1,7,8] and a monocrystal in a diamond-anvil cell.[9] Because pressure applied to powdered material causes stress risers where, say, the sharp edge of one grain touches the face of another, the local pressure at the contact point is always higher than the applied hydrostatic pressure. This effect makes phase transitions studied in powder appear broader with onset at a lower pressure than in bulk. Non-hydrostatic effects also occur in diamond anvil cells. This is a very important issue for this work because we measure substantial, continuous changes well below the critical pressure that we attribute to intrinsic behavior rather than smearing. To circumvent this problem, several monocrystals of ZrW$_2$O$_8$, made using a non-equilibrium technique[10] were used for the measurements reported here. Optically-flat (100) faces were used to measure $C_{11}$ and $C_{44}$, and (110) faces for $C'=(C_{11}-C_{12})/2$ and $C_L=(C_{11}+C_{12}+2C_{44})/2$ for a total of four separate measurements, providing some redundancy because cubic materials possess three independent elastic constants. Sound speed was determined at multiple pressures using an all-digital ultrasonic pulse-echo-overlap method, described elsewhere.[11] Specimen length changes were corrected using the low-pressure elastic constants. Hydrostatic pressure was applied with opposing monocrystal SiC (Moissanite) anvils, Fig. 1, in a pressure cell designed by one of us (Zhao). The pressure-transmitting medium consisted of a mixture of hexagonal boron nitride[12] and fluorinert.[13] A thin acoustically transparent bond between anvil and sample was achieved with cyanoacrylate adhesive. Pressure was determined using ruby fluorescence.[14] All elastic constants were measured in the adiabatic limit.

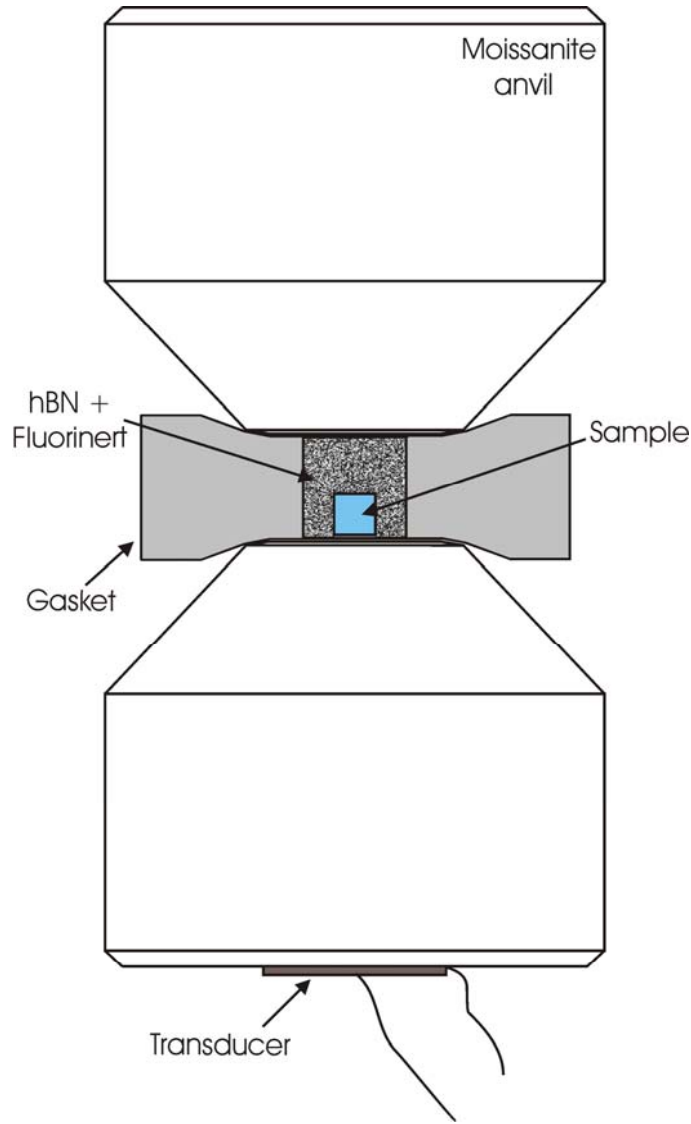

Fig. 1. Schematic of the high-pressure SiC-anvil cell used for determination of elastic constants at high pressure.

Measurements were made from ambient to about 0.71 GPa, spanning the cubic-to-orthorhombic phase transition (0.5 GPa). Elastic constants obtained in the pressure cell at atmospheric pressure agree well with previous resonant ultrasound[15] results,[2] Table I. The relative changes in elastic constants $C_i(P)/C_i(P=0.7 \text{ bar})$ are shown in Fig. 2. Our results show that the critical pressure for the α-γ transition in $ZrW_2O_8$ is 0.5±0.002 GPa, not consistent with previous work.[1,7,8,9] We observe a much sharper phase transition at higher pressure, consistent, as discussed above, with the expected differences between monocrystal and powder studies. Nevertheless, to confirm the critical pressure from elastic constants, we also monitored the Raman spectrum of monocrystal $ZrW_2O_8$ (Fig. 3) as a function of pressure in the same apparatus used to measure elastic constants. We observe only $\alpha$-$ZrW_2O_8$ from 0-0.46 GPa, while at pressures higher than the critical pressure determined from elastic constants (0.5 GPa), only $\gamma$-$ZrW_2O_8$ appears.[9] This confirmation is important in establishing that the pronounced softening of $C_{11}$ and the behavior of the other elastic constants is caused by the underlying physics, and not by non-hydrostatic-pressure-induced smearing of the phase transition.

Table I. Elastic constants at atmospheric pressure and at 300 K.

|  | C11 [GPa] | C44 [GPa] | C' [GPa] | CL [GPa] | B [GPa] |
|---|---|---|---|---|---|
| This study | 127.6 | 26.2 | 38.6 | 114.3 | 75.5 |
| Ref. 2 | 128.4 | 27.4 | 40.4 | 115.4 | 74.5 |

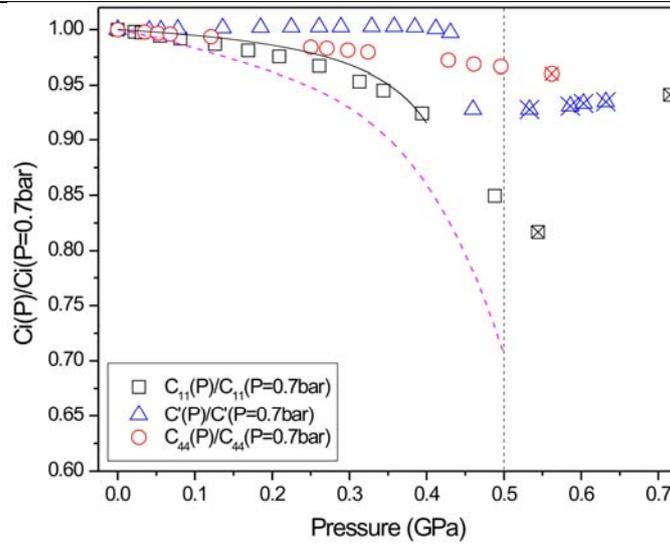

Fig. 2. Relative changes in elastic constants of $ZrW_2O_8$ relative to atmospheric pressure. The bulk modulus B (magenta dashed curve) is computed from a smooth fit to the measured elastic constants (symbols). The black line represents the fit of $C_{11}$ to Eq.(2). Crosses indicate measurements in the $\gamma$-phase of $ZrW_2O_8$. The vertical dotted line is our fit to the phase-transition pressure.

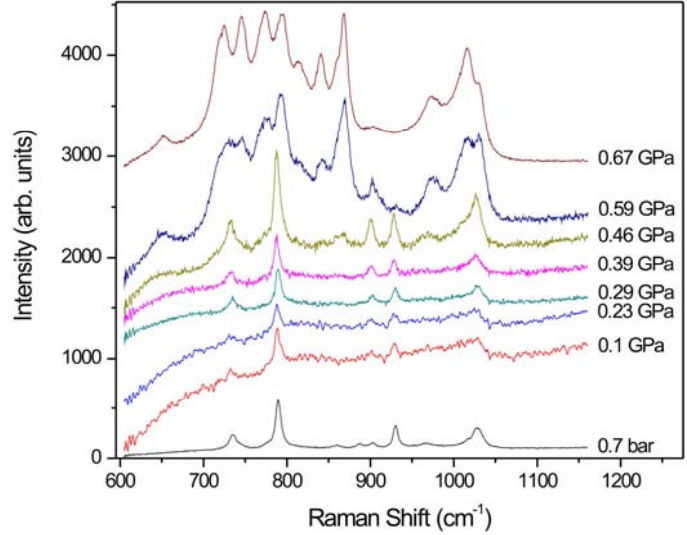

Fig. 3. Raman spectra of $ZrW_2O_8$ as function of pressure for a 600-1150 cm$^{-1}$ interval.

The pressure dependence of $C_{11}$ differs substantially from $C_{44}$, and $C'$. Because the elastic constants were determined in four separate runs, small errors in pressure determination for each loading of the pressure cell, especially at the phase transition where the elastic constants change more rapidly, affect the computed bulk modulus $B$. To obtain a consistent result, $C_{12}$ was calculated from both $C'$ and $C_L$, and then the fits for each were forced to extrapolate to a single critical pressure. Remarkably, $C'$ (shear propagation along [110], polarization along [1$\underline{1}$0]) and $C_{44}$ (shear propagation along [100], polarization along [010]), vary weakly and continuously below the critical pressure, consistent with a cubic solid that should exhibit only small changes in the shear elastic constants as it changes volume because bond angles hardly change. In contrast, all the propagation speeds measured in the high pressure phase ($\gamma$-$ZrW_2O_8$) were lower than in $\alpha$-$ZrW_2O_8$, and increase with pressure, in agreement with results by others[7] (note that the $\gamma$-$ZrW_2O_8$ elastic constants do not exhibit a simple correspondence to those in $\alpha$-$ZrW_2O_8$ because the $\gamma$ (orthorhombic) phase possesses nine independent elastic constants in a tripled unit cell).

One feature of our study seems especially noteworthy in terms of crystal chemistry (structure, bonding, properties): $C_{11}$ (longitudinal propagation along [100], polarization along [100]), and the bulk modulus change excessively, nearly abnormally, as volume decreases, either by pressure or by temperature. To understand why, consider the $ZrW_2O_8$ crystal structure consisting of a three-dimensional-network of corner-linked $WO_4$ tetrahedra and $ZrO_6$ octahedra, a 44-atom unit cell. For simplicity, we consider the diagram shown in Fig. 4, adapted from Simon & Varma.[3] This "constrained lattice" model is different in a very important way from that of Barrera and colleagues,[16] because it has enough "flex" to provide a thermodynamic number of degrees of freedom, rather than a single rigid one. The constrained-lattice model succeeds in explaining the enormous pressure-induced softening of the bulk modulus $B$, and the monocrystal stiffness $C_{11}$, indeed all the dilatation-related elastic constants. To see how, consider the left-side lattice (Fig.4) that corresponds to zero tilting, low temperature, high volume, and high bulk modulus $B$. The right-side corresponds to alternately tilted polyhedra, higher temperature, lower volume, and much lower bulk modulus. Note how the central open space also allows flexing of the rows in many combinations—it is from this that the model derives a large number of degrees of freedom. Applying pressure increases the tilt angle $\theta$ to achieve lower volume. In the $ZrW_2O_8$ crystal structure, two sources of volume-reducing strain are considered. One is the ordinary stiffness of an isolated octahedron, the other arises from tilting and flexing. This second strain mechanism lowers the effective bulk modulus, thus explaining the unusual decrease of $B$ with increasing pressure. An important implication of this second strain is that the magnitude of the thermal expansivity $\beta$ will increase with pressure, opposite the usual decrease that follows from thermodynamics.

It remains to determine the expected pressure dependence of $C_{11}$ based on a constrained-lattice model. A simple approach is to use the solid of Figure 4 to construct a Ginsburg-Landau free energy $F$ that has as the order parameter $\theta$ the angle change from $\pi/2$ between octahedra edges such that

$$F = \frac{1}{2}K\theta^2 + \gamma\theta^4 + P(1-\cos\theta) \qquad (1)$$

where $K$ is the bond-angle stiffness, and $P$ is pressure. Note that we used a simple mechanical picture where $P(1-\cos\theta)$ describes the bending stiffness. For small angles, this is simply a term linear in pressure and quadratic in order parameter. This is the key component—any potential like this will produce a similar result. The minima for this free energy are zero bond angle below a critical pressure, non-zero above. Using this free energy to construct a partition function from which we compute $<\theta^2>$, and noting that the open space of the constrained-lattice model makes $<\theta^2>$ proportional to the decrease in volume $\delta V$, we find that for small changes in the elastic constants, we expect

$$\frac{\delta C_{11}}{C_{11}} \simeq \frac{C_{11}}{P_c}\frac{\delta V}{V} \simeq \frac{C_{11}}{P_c}\frac{Ak_BT}{|P_c-P|} \qquad (2)$$

well below the critical pressure $P_c$, where $A$ is a fitting constant. In Figure 2 we show a fit to Eq. 2. This model not only predicts strong softening of the elastic moduli, but also predicts NTE of the same order as observed and with linear proportionality to temperature.

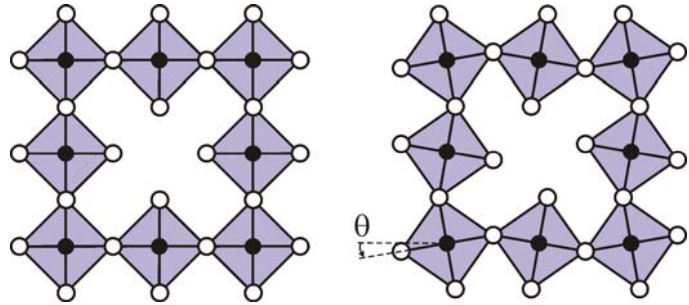

Fig. 4. A simple picture of a framework solid showing how approximately rigid structures that twist can reduce volume because the open space is reduced when twisting occurs. The twisting can be either thermal-motion induced or pressure induced.

The constrained-lattice model, therefore, explains NTE directly, but also predicts quantitatively why $ZrW_2O_8$ softens when its volume is decreased either via NTE and warming, or application of pressure. Essential features of the model include rigid or constrained structures, weak bending stiffness between the structures, and open space that permits a "thermodynamic" number of degrees of freedom. The predictive effectiveness of this model indicates that it is an essential concept in condensed-matter physics, but it also appears attractive in understanding the motions and configuration of more complex systems of interest to biophysics.

Acknowledgements

The National High Magnetic Field Laboratory is supported by the National Science Foundation, Department of Energy and the State of Florida, and Los Alamos National Laboratory Directed Research and Development (LDRD) program X1F7. We would also like to acknowledge the NNSA Campaign funds and LDRD program X1XB, and the DOE grant DE-FG02-02ER45983.